\documentclass[twocolumn]{svjour3}          

\smartqed  

\usepackage{amssymb}
\usepackage{amsmath}
\usepackage{amscd}
\usepackage{latexsym}
\usepackage{graphicx}
\usepackage{color}

\usepackage{cite}

\newcommand{\be}{\begin{equation}}
\newcommand{\ee}{\end{equation}}

\newcommand{\ra}{\rightarrow}
\newcommand{\prt}{\partial}
\newcommand{\al}{\alpha}
\newcommand{\bt}{\beta}
\newcommand{\dlt}{\delta}

\newcommand{\Om}{\Omega}
\newcommand{\om}{\omega}
\newcommand{\gm}{\gamma}

\newcommand{\Gm}{\Gamma}
\newcommand{\lbd}{\lambda}

\newcommand{\dgr}{\dagger}

\newcommand{\bp}{{\bf p}}
\newcommand{\bk}{{\bf k}}
\newcommand{\ba}{{\bf a}}
\newcommand{\bu}{{\bf u}}
\newcommand{\br}{{\bf r}}
\newcommand{\rgl}{\rangle}
\newcommand{\lgl}{\langle}
\newcommand{\cD}{{\cal D}}
\newcommand{\cH}{{\cal H}}

\usepackage{fix-cm}

\usepackage{marvosym}

\newcommand{\envelope}{(\raisebox{-.5pt}{\scalebox{1.45}{\Letter}}\kern-1.7pt)}

\journalname{J Supercond Nov Magn}

\begin{document}

\title{Nanoscale Phase Separation in Ferroelectric Materials
}

\author{V.I. Yukalov \and E.P. Yukalova}

\institute{V.I. Yukalov\envelope \at
Bogolubov Laboratory of Theoretical Physics,
Joint Institute for Nuclear Research, Dubna 141980, Russia   \\              
              \email{yukalov@theor.jinr.ru}          
           \and
           E.P. Yukalova \at
Laboratory of Information Technologies, 
Joint Institute for Nuclear Research, Dubna 141980, Russia}            

\date{Received: date / Accepted: date}

\maketitle

\begin{abstract}
Many materials exhibit nanoscale phase separation, when inside the host thermodynamic 
phase there arise nanosize embryos of another thermodynamic phase. A prominent example 
of this phenomenon is provided by ferroelectric materials. The theoretical description 
of such phase heterogeneous materials is quite challenging, since they are essentially 
nonuniform, the nonuniformity is random, and often they are quasiequilibrium, but not 
absolutely equilibrium. An approach is suggested for the theoretical description of 
phase separated ferroelectrics, consisting of a ferroelectric matrix with nanoscale  
paraelectric inclusions. The properties of the heterophase ferroelectrics are studied. 


\end{abstract}

\section{Features of Nanoscale Separation}

There exists quite a number of materials exhibiting the so-called nanoscale phase 
separation, when in the bulk of one phase there occur nanosize germs of another phase.
The schematic picture of such a heterophase matter is illustrated in Fig. 1. The
typical sizes of the heterophase inclusions $l_f$ are much larger then the 
interparticle distance $a$, but much smaller than the linear size of the sample $L$,
\be
\nonumber
 a \ll l_f \ll L \;  ,
\ee
because of which this phase separation is termed {\it mesoscopic}. Very often, the 
inclusions of the competing phase are not static, but rather dynamic, slightly moving, 
changing their shapes, disappearing and again appearing. This fluctuating nature of 
the germs suggests to call them {\it heterophase fluctuations}. 

\begin{figure}[ht!]
\centerline{
\includegraphics[width=0.65\columnwidth]{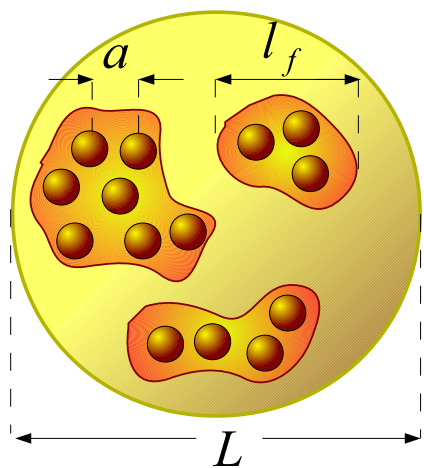} }
\vskip 9pt
\caption{Schematic picture of a sample with mesoscopic phase separation. 
}
\label{fig:Fig.1}
\end{figure}

The features of a fluctuating germ can be compared with the characteristic scales  
typical of condensed matter. Such typical spatial scales, in addition to the 
interparticle distance $a$, are the interaction radius $r_{int}$ and mean-free path
$\lambda_{mfp}$. The latter is estimated as
\be
\nonumber
 \lbd_{mfp} \sim \frac{1}{\rho r_{int}^2} \sim \frac{a^3}{r_{int}^2} \;  ,
\ee
where $\rho \sim a^{-3}$ is average density. The typical velocities are the particle 
velocity and sound velocity
\be
\nonumber
 v \sim s \sim \frac{k_B T_D a}{\hbar} \;  ,
\ee
with $T_D$ being Debye temperature. Then the related temporal scales are the interaction 
time $t_{int}$ and local equilibration time $t_{loc}$,
\be
\nonumber
 t_{int} \sim \frac{r_{int}}{v} \; , \qquad  
t_{loc} \sim \frac{\lbd_{mfp}}{v} \; .
\ee

In condensed matter, the characteristic spatial scales are of order 
\be
\nonumber
 a \sim r_{int} \sim  \lbd_{mfp} \sim 10^{-8} {\rm cm} \;  .
\ee
Debye temperature can be estimated as $T_D \sim 100 K$, which defines the typical 
velocities $v \sim s \sim 10^5$ cm/s and temporal scales
\be
\nonumber
 t_{int} \sim t_{loc} \sim  10^{-13} {\rm s} \; .
\ee

The typical size of a heterophase fluctuation is $l_f \sim 10 - 100$ \AA, hence
the typical temporal scale is 
\be
\nonumber
 t_f \sim \frac{l_f}{v} \sim  10^{-12} \div 10^{-11} {\rm s} \;  .
\ee
This tells us that the heterophase fluctuations are usually also mesoscopic in
time, being between the local-equilibration time and experimental observation time
$t_{exp}$, such that
\be
\nonumber
 t_{loc} \ll t_f \ll t_{exp} \;  .
\ee
   
Such nanoscale phase separation has been observed in high-temperature superconductors 
\cite{Phillips_1,Bianconi_2,Bianconi_3}, ferroelectrics 
\cite{Brookeman_4,Rigamonti_5,Gordon_6,Gordon_7}, around many structural phase 
transitions \cite{Krivoglaz_8,Duvall_9,Bruce_10,Bakai_11}, in macromolecular assemblies
\cite{Yukalov_12}, and at liquid-glass transitions \cite{Bakai_13}. More references can 
be found in review articles \cite{Yukalov_14,Yukalov_15,Yukalov_16}. In the present paper, 
we concentrate on the study of heterophase ferroelectrics, in which inside ferroelectric 
phase there exist paraelectric nanoscale bubbles.

\section{Phase Separation Description}

Mathematically, the general situation, when the sample is phase separated, can be 
described in the following way \cite{Yukalov_14,Yukalov_15,Yukalov_16}. Let the phases
be enumerated by the index $\nu = 1,2,\ldots$. At a given snapshot, the system space
$\mathbb{V}$ is separated into subspaces $\mathbb{V}_\nu$, occupied by the related
phases and forming an orthogonal covering $\{\mathbb{V}_\nu\}$, such that  
\be
\label{1}
 \mathbb{V} = \bigcup_\nu \mathbb{V}_\nu \;  .
\ee
The corresponding subvolumes sum to the total system volume
\be
\label{2}
 V = \sum_\nu V_\nu \qquad ( V_\nu \equiv {\rm mes} \mathbb{V}_\nu ) \;  .
\ee
The separation of thermodynamic phases can be done by means of an equimolecular
surface \cite{Gibbs_17,Gibbs_18}, when the total number of particles is the sum
\be
\label{3}
 N = \sum_\nu N_\nu \;  .
\ee

The spatial location of the phases is fixed by the manifold indicator functions
\begin{eqnarray}
\label{4}
\xi_\nu(\br) = \left \{ \begin{array}{ll}
1 , ~ & ~ \br \in \mathbb{V}_\nu \\
0 , ~ & ~ \br \not\in \mathbb{V}_\nu
\end{array}  . \right.
\end{eqnarray}
The overall phase configuration is described by the collection
\be
\label{5}
\xi \equiv \{ \xi_\nu(\br) : \; \nu = 1,2,\ldots; \; \br \in \mathbb{V} \} \; .
\ee
  
Under a given phase configuration, the statistical operator $\hat{\rho}(\xi)$ can be 
found from the principle of minimal information, keeping in mind the normalization
condition
\be
\label{6}
 {\rm Tr} \int \hat\rho(\xi) \;\cD\xi = 1 \;  ,   
\ee
where the trace operation is over the quantum degrees of freedom, while the functional 
integration is over the manifold indicator functions that paly the role of random 
variables. Also, there is the definition of the internal energy
\be
\label{7}
  E =  {\rm Tr} \int \hat\rho(\xi) H(\xi) \; \cD\xi \;  ,
\ee
which is the average of a Hamiltonian $H(\xi)$. The information functional has the 
form
\be
\nonumber
I[ \;\hat\rho(\xi)\; ] = 
{\rm Tr} \int \hat\rho(\xi) \ln \; \frac{\hat\rho(\xi)}{\rho_0(\xi)} \; \cD\xi +
\ee
\be
\nonumber
+
\al \left [ {\rm Tr} \int \hat\rho(\xi)\; \cD\xi - 1  \right ] +
\ee
\be
\label{8}
+ 
\bt \left [ {\rm Tr} \int \hat\rho(\xi) H(\xi)\; \cD\xi - E \right ] \; ,
\ee
where the first term is the Kullback-Leibler information \cite{Kullback_19,Kullback_20}, 
$\alpha$ and $\beta$ are the Lagrange multipliers guaranteeing the validity of conditions 
(\ref{6}) and (\ref{7}), and $\hat{\rho}_0$ is a prior statistical operator taking into 
account additional prior information on the system, when it is available. If no additional 
apriori information is provided, the prior statistical operator is proportional to unity
operator. 

Minimizing the information functional yields the statistical operator
\be
\label{9}
 \hat\rho(\xi) = \frac{\hat\rho_0(\xi) \exp\{ -\bt H(\xi) \} }
{{\rm Tr}\hat\rho_0(\xi) \exp\{ -\bt H(\xi) \} } \; .
\ee
And the system thermodynamic potential, say free energy, writes as 
\be
\label{10}
 F = - T \ln {\rm Tr} \int \hat\rho_0(\xi) \exp\{ -\bt H(\xi) \} \; \cD\xi \; ,
\ee
where $\beta T = 1$. All thermodynamic properties of the heterophase system can be 
defined if the thermodynamic potential can be calculated. 

In (\ref{10}) and in what follows, we set the Planck and Boltzmann constants as unity.

\section{Averaging over Phase Configurations}

The averaging over phase configurations is represented by the functional integration 
over the manifold indicator functions. In order to accomplish calculations involving this
integration, it is necessary to explicitly define the corresponding functional measure.

Let us introduce for each subspace $\mathbb{V}_\nu$ an orthogonal subcovering
$\{ \mathbb{V}_{\nu i}\}$, such that
\be
\label{11}
 \mathbb{V}_\nu = \bigcup_{i=1}^{n_\nu} \mathbb{V}_{\nu i} \;  .
\ee
Then the manifold indicator (\ref{4}) can be written as the sum
\be
\label{12}
 \xi_\nu(\br) = \sum_{i=1}^{n_\nu} \xi_{\nu i} (\br - \ba_{\nu i} ) \;  ,
\ee
with the submanifold indicators
\begin{eqnarray}
\label{13}
\xi_{\nu i}(\br) = \left \{ \begin{array}{ll}
1 , ~ & ~ \br \in \mathbb{V}_{\nu i} \\
0 , ~ & ~ \br \not\in \mathbb{V}_{\nu i}
\end{array}  . \right.
\end{eqnarray}

Snapshot phase weights are given by the integrals
\be
\label{14}
 \xi_\nu \equiv \frac{1}{V} \int \xi_\nu(\br) \; d\br = 
\int \xi_\nu(\br)\; \cD\xi \;  , 
\ee
satisfying the normalization conditions
\be
\label{15}
 \sum_\nu \xi_\nu = 1 \; , \qquad 0 \leq \xi_\nu \leq 1 \;  ,
\ee
which defines the set $\{\xi_\nu\}$ as a probability measure. The differential measure
over the set of all possible configurations is 
\be
\label{16}
 \cD\xi = \dlt\left ( \sum_\nu \xi_\nu -1 \right ) \prod_\nu d\xi_\nu \;
\prod_\nu \prod_{i=1}^{n_\nu} \frac{d\ba_{\nu i} }{V} \;  ,
\ee
under the asymptotic condition $n_\nu \ra \infty$.  
 
Defining an effective Hamiltonian $\tilde{H}$ by the relation
\be
\label{17}
\exp ( - \bt \widetilde H ) = \int \hat\rho_0(\xi) \exp\{ -\bt H(\xi) \}\; \cD\xi
\ee
makes it straightforward to rewrite the thermodynamic potential (\ref{10}) in the simple
form
\be
\label{18}
 F = - T \ln {\rm Tr} e^{-\bt \widetilde H} \;  ,
\ee 
where $\beta T = 1$. This potential depends on the geometric phase probabilities $w_\nu$ 
that are the minimizers of the thermodynamic potential,
\be
\label{19}
 F = F (\{ w_\nu\} ) = {\rm abs}\min_{ \{\xi_\nu\} } F (\{ \xi_\nu\} ) .
\ee
The phase probabilities satisfy the normalization condition
\be
\label{20}
 \sum_\nu w_\nu = 1 \; , \qquad 0 \leq w_\nu \leq 1 \;  .
\ee
Thus, defining the effective Hamiltonian by relation (\ref{17}), we reduce the problem 
for a nonuniform system to the consideration of an effective uniform system with an
effective Hamiltonian $\tilde{H}$.

\section{Model of Heterophase Ferroelectric}

Now we apply the above techniques for considering a heterophase ferroelectric that
consists of a ferroelectric matrix with nanoscopic inclusions of paraelectric bubbles.
The latter are randomly distributed in the sample volume, with no prior information
on their locations being available. Recall that heterophase fluctuations are known 
to exist in many ferroelectrics \cite{Brookeman_4,Rigamonti_5,Gordon_6,Gordon_7,Yukalov_21},
such as, e.g., HCl and HCl-DCl. For concreteness, we consider here the ferroelectrics of
the order-disorder KDP type \cite{Blinc_22}, although a similar consideration can be 
realized for the ferroelectrics of displacement type. 

The derivation of the Hamiltonian for a ferroelectric of the KDP type is similar to that
for a double-well optical lattice filled by cold atoms \cite{Yukalov_23,Yukalov_24,Yukalov_25}.
Accomplishing the averaging over phase configurations, as described above, we come to the
effective Hamiltonian
\be
\label{21}
 \widetilde H = H_1 \bigoplus H_2 \;  ,
\ee
with the phase components
\be
\nonumber
H_\nu = w_\nu \sum_j ( K_j - \Om S_j^x - B_0 S_j^z ) +
\ee
\be
\label{22}
 +
w_\nu^2 \sum_{i\neq j} \left ( \frac{1}{2} \; A_{ij} + B_{ij} S_i^x S_j^x -
I_{ij} S_i^z S_j^z \right ) \;   .
\ee
Here $K_j$ is a single-site energy (a matrix element of kinetic energy), $\Omega$ is
tunneling frequency, $B_0$ is a strain field caused by external forces, $A_{ij}$ is a 
matrix element of direct particle interactions, and $B_{ij}$ and $I_{ij}$ are matrix 
elements of exchange interactions. The quasi-spin operators have the following meaning.
The operator $S_j^x$ describes particle tunneling between the wells of a double well 
potential in a $j$-th lattice site. The operator $S_j^y$ corresponds to the Josephson 
current between the wells. And the operator $S_j^z$ characterizes the particle imbalance 
of the wells. The factors $w_\nu$ are the phase probabilities defined in the previous 
sections. Each term $H_\nu$ acts on a weighted Hilbert space $\mathcal{H}_\nu$, with 
the total Hamiltonian (\ref{21}) acting on the fiber space
\be
\label{23}
 \cH = \cH_1 \bigotimes \cH_2 \;  .
\ee
The value of $B_{ij}$ is usually much smaller than that of $I_{ij}$,
\be
\label{24}
|\; B_{ij} \; | \ll |\; I_{ij} \; | \;  ,
\ee
because of which it can be omitted.      
  
Since there are two thermodynamic phases, there exist two order parameters
\be
\label{25}
 s_\nu \equiv \frac{2}{N} \sum_j \; \lgl \; S_j^z \; \rgl_\nu \qquad
( \nu = 1,2 ) \;  ,
\ee
describing an average particle imbalance in each phase, where
\be
\nonumber
\lgl \; S_j^z \; \rgl_\nu = 
\frac{{\rm Tr}_{\cH_\nu} S_j^z\exp(-\bt H_\nu)}{{\rm Tr}_{\cH_\nu}\exp(-\bt H_\nu)} \; .
\ee
Ferroelectric phase enjoys a larger order parameter,
\be
\label{26}
 s_1 > s_2 \;  .
\ee
When there is no external strain, then
\be
\label{27}
 s_2 = 0 \qquad ( B_0 = 0 ) \;  ,
\ee
which implies the absence of polarization in paraelectric phase, when there are 
no external fields \cite{Yukalov_21,Yukalov_26}. 

To realize explicit calculations, we need to invoke a decoupling for the quasi-spin 
operators. Here we resort to the Kirkwood decoupling \cite{Kirkwood_27} having he form
\be
\nonumber
S_i^\al S_j^\bt = g_{ij}^\nu \left [ \; \lgl \; S_i^\al \; \rgl_\nu \; S_j^\bt \; + 
\right.
\ee
\be
\label{28}
  \left. + \;
 S_i^\al \; \lgl \; S_j^\bt \; \rgl_\nu - 
\lgl \; S_i^\al \; \rgl_\nu \; \lgl \; S_j^\bt \; \rgl_\nu \; \right ] \; ,
\ee
in which $g_{ij}^\nu$ characterizes particle correlations. In ferroelectric phase,
the correlations are long-ranged, while in paraelectric phase, they are short-ranged.   
 
For what follows, we need the notation for the correlation parameter
\be
\label{29}
g_\nu \equiv \frac{\sum_{i\neq j} I_{ij} g_{ij}^\nu}{\sum_{i\neq j} I_{ij} }
\ee
and also the notations
\be
\label{30}
 u \equiv \frac{A}{J} \; , \qquad h \equiv \frac{B_0}{J} \;  ,
\ee
where
\be
\nonumber
 A \equiv \frac{1}{N} \sum_{i\neq j} A_{ij} \; , \qquad
 J \equiv \frac{1}{N} \sum_{i\neq j} I_{ij} \;  .
\ee
The dimensionless parameter $u$ characterizes direct disordering interactions, as 
compared to exchange ordering interactions, while the parameter $h$ defines the 
strength of external strain.

\section{Phonon degrees of freedom}

Phonons play an important role in ferroelectrics. The related collective excitations
can be introduced in the way that has been used for defining phonons in quantum
crystals \cite{Guyer_28} or in optical lattices \cite{Yukalov_29}. 

The interaction terms in Hamiltonian (\ref{22}) are assumed to depend on the locations
of particles as $A({\bf r}_i - {\bf r}_j)$ and $I({\bf r}_i - {\bf r}_j)$. Each particle
location is represented as
\be
\label{31}
 \br_j = \ba_j + \bu_j \;  ,
\ee
where
\be
\label{32}
 \ba_j \equiv \lgl \br_j \rgl_\nu \; , \qquad 
\lgl \bu_j \rgl_\nu = 0 \;  .
\ee
Then the difference between the locations of two particles writes as
\be
\nonumber
\br_{ij} \equiv \br_i - \br_j =  \ba_{ij} + \bu_{ij} \; , 
\ee
\be
\nonumber
\ba_{ij} \equiv \ba_i - \ba_j \; , \qquad \bu_{ij} \equiv \bu_i - \bu_j \; .
\ee

Keeping in mind that the deviations of particles from their lattice sites are small,
the interaction terms are expanded in powers of the deviations, limiting ourselves 
by the second-order powers, which gives
\be
\nonumber
A(\br_{ij}) \cong A_{ij} + \sum_\al A_{ij}^\al u_{ij}^\al \; - \; 
\frac{1}{2} \sum_{\al\bt} A_{ij}^{\al\bt} u_{ij}^\al u_{ij}^\bt \;  ,
\ee
where
\be
\nonumber
 A_{ij} \equiv A(\ba_{ij}) \; , \qquad 
A_{ij}^\al \equiv \frac{\prt A_{ij}}{\prt a_i^\al} \; , \qquad
A_{ij}^{\al\bt} \equiv \frac{\prt^2 A_{ij}}{\prt a_i^\al\prt a_j^\bt} \; .
\ee
Then Hamiltonian (\ref{22}) transforms into
\be
\nonumber
H_\nu = w_\nu \sum_j ( K_j - \Om S_j^x - B_0 S_j^x ) \; + 
\ee
\be
\nonumber
+ \;
\frac{1}{2} w_\nu^2 \sum_{i\neq j} \left ( 
A_{ij} - \; \frac{1}{2} \sum_{\al\bt} A_{ij}^{\al\bt}u_{ij}^\al u_{ij}^\bt 
\right ) -
\ee
\be
\label{33}
 - \; w_\nu^2 \sum_{i\neq j} \left ( I_{ij} + \sum_\al I_{ij}^\al u_{ij}^\al 
- \; \frac{1}{2} \sum_{\al\bt} I_{ij}^{\al\bt}u_{ij}^\al u_{ij}^\bt
\right ) S_{ij}^z \; ,
\ee
with the notation
\be
\label{34}
S_{ij}^z \equiv S_i^z S_j^z \; .
\ee
   
Phonon and quasi-spin variables are decoupled so that to yield
\be
\label{35}
 \lgl \; u_{ij}^\al u_{ij}^\bt S_{ij}^\gm \; \rgl_\nu  =  
\lgl \; u_{ij}^\al u_{ij}^\bt \; \rgl_\nu \; \lgl \; S_{ij}^\gm \; \rgl_\nu \; .
\ee

The phonon spectrum is defined by the eigenproblem
\be
\label{36}
 \frac{w_\nu}{m} \sum_{j(\neq i)} \sum_\bt \Phi_{ij}^{\al\bt} e^{i\bk\cdot\ba_{ij}}
e^\bt_{ks} = \om_{ks}^2 e_{ks}^\al \;  ,
\ee
with the renormalized matrix
\be
\label{37}
 \Phi_{ij}^{\al\bt} \equiv A_{ij}^{\al\bt} - 
2 I_{ij}^{\al\bt} \lgl S_{ij}^z \rgl_\nu \; .
\ee
Here ${\bf e}_{ks}$ is a polarization vector, with $s$ being the polarization index. 

The phonon destruction and creation operators are introduced by the relations
\be
\nonumber
\bp_j = - \; \frac{i}{\sqrt{2N}} \sum_{ks} \sqrt{m\om_{ks}}\; {\bf e}_{ks}
\left ( b_{ks} - b_{-ks}^\dgr \right ) e^{i\bk \cdot \ba_j} \; ,
\ee
\be
\label{38}
 \bu_j = {\bf v}_j + \; 
\frac{1}{\sqrt{2N}} \sum_{ks} \frac{ {\bf e}_{ks} }{ \sqrt{ m\om_{ks} } }\; 
\left ( b_{ks} + b_{-ks}^\dgr \right ) e^{i\bk \cdot \ba_j} \;  .
\ee
Notice that the second transformation is nonuniform, which is necessary for getting 
rid of the terms linear in the operators $b_{ks}$, as is discussed in \cite{Yukalov_30}. 
In the present case,
\be
\nonumber
 v_f^\al = -\; \frac{w_\nu}{2N} \sum_{i\neq j} 
\sum_\bt \gm_{fj}^{\al\bt} I_{ij}^\bt S_{ij}^z \;  ,
\ee
with
\be
\nonumber
 \gm_{jf}^{\al\bt} \equiv 4 \sum_{ks} \frac{e_{ks}^\al e_{ks}^\bt}{m\om_{ks}^2} \;
e^{i\bk \cdot\ba_{jf} } \;  .
\ee

Then Hamiltonian (\ref{33}) reduces to the sum
\be
\label{39}
 H_\nu = E_\nu + H_\nu^{ph} + H_\nu^{ps} + H_\nu^{ind} \;  .
\ee
The first term here is 
\be
\label{40}
 E_\nu = \frac{1}{2}\; w_\nu^2 NA - 
w_\nu^2 \sum_{i\neq j} \sum_{\al\bt} I_{ij}^{\al\bt} \lgl \; S_{ij}^z \; \rgl_\nu
\; \lgl \; u_j^\al u_j^\bt \; \rgl_\nu \; .
\ee
The second term is the phonon Hamiltonian
\be
\label{41}
 H_\nu^{ph} = w_\nu \sum_{ks} \om_{ks} \left ( b_{ks}^\dgr b_{ks} + \frac{1}{2}
\right ) \;  .
\ee
The third term is the pseudospin Hamiltonian
\be
\label{42}
   H_\nu^{ps} = - w_\nu \sum_j \left ( \Om S_j^x + B_0 S_j^z \right ) -
 w_\nu^2 \sum_{i\neq j} \widetilde I_{ij} S_{ij}^z \;  ,
\ee
with the renormalized interaction
\be
\label{43}
  \widetilde I_{ij} \equiv I_{ij} - \sum_{\al\bt} I_{ij}^{\al\bt}
\lgl \; u_j^\al u_j^\bt \; \rgl_\nu \; .
\ee
And the last term in (\ref{39}) is the four-spin Hamiltonian induced by the 
interactions of quasi-spins through particle oscillations,  
\be
\label{44}
 H_\nu^{ind} = -\; \frac{w_\nu^3}{N} 
\sum_{i\neq j} \sum_{f\neq g} \Gm_{ijfg} S_{ij}^z S_{fg}^z \; ,
\ee
with the vertex
\be
\nonumber
 \Gm_{ijfg} \equiv \sum_{\al\bt} I_{ij}^\al \gm_{jf}^{\al\bt} I_{fg}^\bt \;  .
\ee
It is possible to show that the induced shift ${\bf v}_j$ as well as the induced
Hamiltonian are small \cite{Yukalov_16}. In the long-wave approximation, we have
\be
\nonumber
 {\bf v}_j \cong 0 \; , \qquad H_\nu^{ind} \cong 0 \;  .
\ee
Therefore the deviation-deviation correlation function becomes
\be
\label{45}
 \lgl u_i^\al u_j^\bt \rgl_\nu = 
\frac{\dlt_{ij}}{2N} \sum_{ks} \frac{e_{ks}^\al e_{ks}^\bt}{m\om_{ks}} \; 
\coth \left ( \frac{w_\nu \om_{ks}}{2T} \right ) \;  .
\ee

\section{Properties of Heterophase Ferroelectrics}

To study the properties of the heterophase ferroelectric, we accomplish numerical
calculations for the model of Sec. 4. The quasi-spin variables are treated in the 
Kirkwood approximation (\ref{28}), with the correlation parameters $g_1 = 1$ and
$g_2 \ll 1$. Also we take into account that $\Omega \ll J$. The phase probability
$w_\nu$ is defined as the minimizer of the free energy (\ref{18}), according to
conditions (\ref{19}) and (\ref{20}).

\begin{figure}[ht!]
\centerline{
\includegraphics[width=0.85\columnwidth]{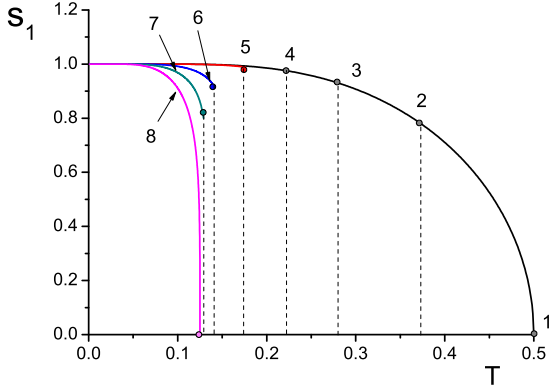} }
\caption{Ferroelectric order parameter $s_1$, in the absence of strain, 
$h = 0$, as a function of temperature $T$, in units of $J$, for different 
disorder parameters: (1) $u = 0$; (2) $u = 0.1$; (3) $u = 0.3$; (4) $u = 0.4$; 
(5) $u = 0.51$; (6) $u = 0.75$; (7) $u = 1$; (8) $u = 1.5$. The points mark 
the temperatures of ferroelectric-paraelectric phase transitions.
}
\label{fig:Fig.2}
\end{figure}

Figure 2 shows the temperature behavior of the ferroelectric order parameter $s_1$ 
in the absence of external strain, $h = 0$, for different disorder parameters $u$.
Temperature is measured in units of the exchange interaction strength $J$. The 
existence of the mesoscopic paraelectric germs inside the ferroelectric matrix makes 
the phase transition ferroelectric-paraelectric of first order for the disorder 
parameters in the interval $0 < u < 3/2$. The phase transition temperature is in 
the range $0.125 < T_c < 0.5$.

Figure 3 demonstrates the influence of the external strain $h$ for the fixed disorder
parameter $u = 0.3$. The phase transition for this $u$ is of first order, but the 
strain makes $s_1$ nonzero above the transition point.

\begin{figure}[ht!]
\centerline{
\includegraphics[width=0.85\columnwidth]{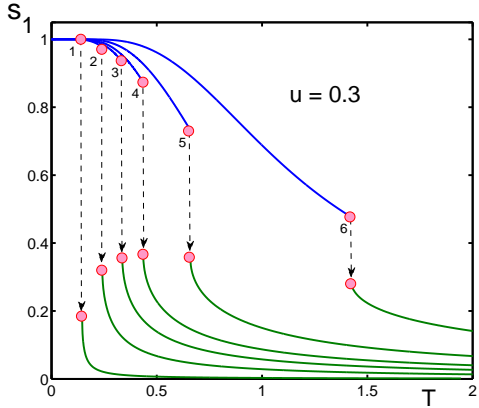} }
\caption{Ferroelectric order parameter $s_1$, as a function of temperature, 
in units of $J$, for fixed disorder parameter $u = 0.3$ and varying external 
strain: (1) $h = 0.01$; (2) $h = 0.1$; (3) $h = 0.2$; (4) $h = 0.3$; 
(5) $h = 0.5$; (6) $h = 1$. The corresponding first-order transition temperatures 
are: (1) $T_0 = 0.145$; (2) $T_0 = 0.241$; (3) $T_0 = 0.336$; (4) $T_0 = 0.436$; 
(5) $T_0 = 0.660$; (6) $T_0 = 1.422$.
}
\label{fig:Fig.3}
\end{figure}

The sound velocity for a heterogeneous matter is expected to be lower than that of
the pure phase. The isotropic part of the sound velocity writes as
\be
\label{46}
 s = \sum_\nu w_\nu s_\nu \;  ,
\ee
where
\be
\label{47}
s_\nu \equiv \lim_{k\ra 0} \; \frac{\om_k}{k} \qquad 
\left ( \om_k^2 = \frac{1}{3} \sum_{s=1}^3 \om_{ks}^2 \right ) \;   .
\ee
From Sec. 5, we have
\be
\label{48}
 s_\nu = \sqrt{w_\nu}\; c_\nu \qquad 
\left ( c_\nu \equiv \sqrt{ \frac{D_\nu}{2m} } \right ) \;  ,
\ee
where $D_\nu$ denotes the nearest-neighbor part of the dynamic matrix
\be
\nonumber
 D_{ij}^\nu \equiv - \; \frac{1}{3} \sum_{\al=1}^3 \Phi_{ij}^{\al\al} \;  .
\ee
As a result, we get
\be
\label{49}
  s = w_1^{3/2} c_1 + w_2^{3/2} c_2 \; .
\ee

Since $w_\nu < 1$, and taking into account that $c_1 \approx c_2 \equiv c$, 
it is seen that $s < c$. The maximal attenuation of the sound velocity occurs at
the transition point, where $w_1 \approx 0.5$. Then the relative decrease of the 
sound velocity, caused by heterophase fluctuations, is 
$\delta s \equiv (s - c)/c \approx - 0. 293$.   

The mesoscopic heterophase inclusions increase the mean-square deviation of particles
\be
\label{50}
  r_\nu^2 \equiv \sum_\al \lgl u_i^\al u_i^\al \rgl_\nu = 
\frac{1}{2N} \sum_{ks} \frac{1}{m\om_{ks}} \; \coth \frac{w_\nu \om_{ks}}{2T} \;  .
\ee
Thus, in the Debye approximation, when
\be
\label{51}
 \om_{ks} \equiv \om_k = s_\nu k \; \Theta(k_D - k) \;  ,
\ee
with the Debye radius defined by the expression $k_D^3 = 6 \pi^2 \rho$, the mean-square 
deviation is
\be
\label{52}
r_\nu^2 = \frac{18T^2 w_\nu}{m\Theta_\nu^3} 
\int_0^{\Theta_\nu/2T} x\coth x \; dx \; ,
\ee
where the effective Debye temperature is
\be
\label{53}
 \Theta_\nu \equiv w_\nu s_\nu k_D = w_\nu^{3/2} T_{\nu D} \qquad
(T_{\nu D} \equiv c_\nu k_D ) \;  .
\ee
At relatively low or high temperature, one has
\begin{eqnarray}
\nonumber
r_\nu^2 \simeq \left \{ \begin{array}{ll}
9w_\nu/ 4m \Theta_\nu , ~ & ~ T \ll \Theta_\nu \\
\\
9w_\nu T/ m\Theta_\nu^2 , ~ & ~ T \gg \Theta_\nu
\end{array} . \right.
\end{eqnarray}
This shows that the $\nu$-th phase can be treated as localized only when its
weight is sufficiently large, so that $r_\nu > a$. Otherwise, the particles cannot 
be localized, experiencing strong diffusion \cite{Mehrer_31}, hence cannot form 
localized heterophase germs. 

The increase of the mean-square deviation influences the value of the Debye-Waller
factor that, for a heterophase system, has the form
\be
\label{54}
f_{DW} = \sum_\nu w_\nu f_\nu \;   ,
\ee
where
\be
\nonumber
f_\nu = \exp \left ( - \;\frac{1}{3} \; k_0^2 r_\nu^2 \right ) \;   .
\ee
The same expression is valid for the M\"{o}ssbauer effect probability \cite{Yukalov_32}.

The latter, with the notation for the recoil energy $E_R \equiv k_0^2/2m$ reads as
\be
\label{55}
f_\nu = \exp \left ( - \;\frac{2}{3} \; m E_R  r_\nu^2 \right ) \;   .
\ee
The typical behavior of the Debye-Waller factor (\ref{54}) is shown in Fig. 4. At 
the transition temperature, the factor exhibits the so-called {\it cusp-shaped anomaly}.
The corresponding relative sagging is about $30\%$, as compared to the value just 
above $T_c$.

\begin{figure}[ht!]
\vspace{9pt}
\centerline{
\includegraphics[width=0.85\columnwidth]{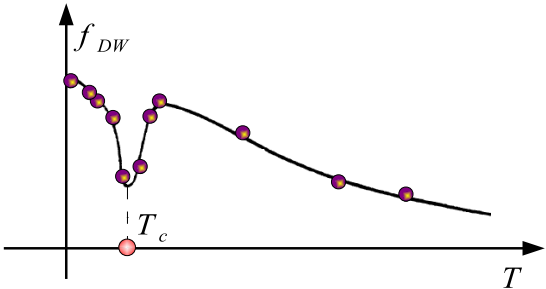} }
\caption{Debye-Waller factor as a function of temperature, exhibiting the typical 
cusp-shaped anomaly at $T_c$, caused by the arising heterophase fluctuations.
}
\label{fig:Fig.4}
\end{figure}

This behavior is in good agreement with the Debye-Waller factor 
(or Lamb-M\"{o}ssbauer factor) of many ferroelectrics, displaying the cusp-shaped 
anomaly at $T_c$. First, it has been observed for BaTiO$_3$ and PbTiO$_3$ at a weak 
first-order phase transition \cite{Bhide_33,Bhide_34,Bhide_35} and later for many 
other ferroelectrics and antiferroelectrics, as well as at magnetic transitions,
structural transitions, and high-temperature superconducting transitions
\cite{Bhide_36,Owens_37,Bishop_38,Egami_39,Muller_40,Sirdeshmukh_41}. Initially,
one tried to connect such cusp-shaped anomalies with the existence of soft modes.
However, by accurate microscopic treatments, it has been proved that soft modes
are able to account for only about $1 \%$ of change in the Debye-Waller factor and 
cannot be related to such a large anomaly as a $30 \%$ sagging of the factor 
\cite{Meissner_42,Binder_43,Yukalov_44}. But the cusp-shaped anomaly can be explained
by the presence of heterophase fluctuations, as is shown above.

\section{Conclusion}

We have presented a theory of ferroelectrics, inside which there exist fluctuating
germs of paraelectric phase. The appearance of such heterophase fluctuations essentially
influences the properties of the matter. For instance, the sound velocity decreases,
the mean-square deviation increases, and the Debye-Waller factor experiences a cusp-shaped
anomaly at the point of a phase transition. The described cusp-shaped anomaly is in good 
agreement with experiments. The presented method of describing heterophase fluctuations 
can be employed for other types of condensed matter, e.g., for high temperature 
superconductors \cite{Yukalov_45}.

\begin{acknowledgements}
The authors acknowledge financial support from the RFBR (grant $\# 14-02-00723$).
\end{acknowledgements}

%
%

\end{document}